\documentclass[10pt]{article}     % you may use here any options admissible 
\usepackage[pdftex]{graphicx}
\DeclareGraphicsExtensions{.eps,.jpeg,.pdf} %frv
\usepackage{amsmath}
\usepackage{epstopdf}
\usepackage[margin=1.4in]{geometry}   %frv
\begin{document}

\title{\bf Passive and active electromagnetic stabilization of free-surface 
liquid metal flows}	%Paper title
\author{S.M.H.~Mirhoseini, R.R.~Diaz-Pacheco, F.A.~Volpe$^*$\\
{\sl Department of Applied Physics and Applied Mathematics}\\
{\sl Columbia University, New York, NY 10027, USA}\\
* Corresponding author: fvolpe@columbia.edu}

%----------------------- End of Preamble

%\begin{document}

%----------------------- Head
\maketitle
%\dedication{...}			%if any
\begin{abstract}%
Free-surface liquid metal flows tend to be uneven due to instabilities 
and other effects. 
Some applications, however, require constant, uniform liquid metal thickness. 
This is for example the case of liquid walls in nuclear fusion reactors. 
With this motivation, 
here we present experimental results on the stabilization of 
a free-surface flow of Galinstan. 
The flow was sustained by an electromagnetic induction 
pump featuring rotating permanent magnets. Evidence is reported of the flowing 
Galinstan layer becoming flatter in the presence of a sufficiently strong 
magnetic field, either alone (passive stabilization) or in combination 
with an electrical current passing through the liquid metal 
(active stabilization). 
The results are interpreted in terms of an effective viscosity and effective 
gravity, respectively. 
%\classification{PACS number(s)}
\end{abstract}
%----------------------- End of Head     

%----------------------- Body       
\section*{Introduction.}
\label{sec:intro} 
Free-surface flows of Lithium and other liquid metals are considered 
a promising plasma-facing ``first wall'' for a nuclear fusion reactor: 
liquid walls could protect the underlying solid walls from high heat 
and neutron fluxes \cite{Moir,ref5}, and have been shown to lead to cleaner 
\cite{ref7,MajeskiCDX,MajeskiNF,Nygren} and 
more stable plasmas \cite{ref8,Carlos}. 

However, free-surface liquid metal (LM)  
layers will tend to be uneven \cite{ref9} as a result 
of non-uniform force fields, liquid metal instabilities and 
turbulence \cite{ref11}. 
Uneven LM surfaces could enter in contact with the plasma, contaminate it, cool 
it, and possibly disrupt it, or they might expose the underlying solid
wall to damage by heat and neutrons, and expose the plasma to a 
material with less benign erosion and wall recycling properties. 

This motivated our investigation of how to stabilize free-surface liquid 
metal flows. As reported in the present article, 
this could be achieved ``passively'', thanks to the strong 
magnetic field naturally present in a magnetically confined fusion reactor, or 
``actively'', by applying an electrical current in the liquid metal. 

Note that, in addition to the deliberately applied current-pattern, 
other currents will naturally be induced in the liquid metal, e.g.~due to 
rotating modes in the plasma, and additional current fluctuations could 
result from turbulence. These additional induced or fluctuating 
currents will be non-uniform and will change with 
time. As a result, the ${\bf j}\times{\bf B}$ force-density will change, 
and adjustments of the applied current-pattern might be required, in feedback 
with multi-point measurements of liquid metal thickness. This is the subject of 
separate works \cite{ref11,Mirho2}. 

In this paper we start by discussing potential applications of 
stabilized, free-surface LM flows outside of fusion (Sec.\ref{SecApplic}). 
In Sec.\ref{SecSetup} we describe the experimental setup, including 
the electromagnetic induction pump and a tiltable ``tile'' where the otherwise 
ducted flow is ``free-surface'' and exposed to a magnetic field ${\bf B}$, 
but not to a plasma, yet. Currents of density ${\bf j}$ are also applied. 
Passive stabilization by ${\bf B}$ only and active 
stabilization by ${\bf j}\times {\bf B}$
are reported respectively in Sec.\ref{SecPass} and \ref{SecAct}, along 
with their interpretations.

%==========================================================================
\section{Motivation: free-surface liquid metal flows.}     \label{SecApplic}
%==========================================================================

Beside fusion-related uses, LM surface control has other
scientific and industrial applications. 

Controlling liquid metal flows can be of
interest in metallurgy, for example to control a casting flow, to
remove undesired droplets in laser- or plasma-cutting of metals or, on
the contrary, to control and recover spills of precious materials, for
economic reasons, or spills of toxic or unsafe materials, in order to
protect public health and the environment.  

Another area is
furnaces. Conventional refractory materials are solid, but no solid
material exists, at atmospheric pressure, above the melting point
(4215 K) of Tantalum hafnium carbide. However, several
metals (tantalum, tungsten, rhenium) are liquid in the 4200-6200 K
temperature range. A good control of these metals in their liquid
phase would allow coating the inner walls of a furnace and operating
it at unprecedented temperatures. This would open the way to
high-temperature chemistry in normally inaccessible regimes where
certain undesired reactions are disfavored and certain elusive
reactions are favored \cite{ref12}, from which new processes and
materials could result. 

At the opposite end of the spectrum of “heavy industry” are two
emerging “high-tech” applications of LMs, which could benefit from 
passive, active or feedback stabilization. One technology is 3D metal printing
\cite{ref13}. The other makes use of Galinstan to realize flexible
electronics, tunable metamaterials and microfluidic
devices \cite{ref13,ref14,Hawaii}. Both would benefit highly from flow control.

Finally, LMs are used to generate large concave reflective surfaces in
some optical telescopes \cite{ref15}:
large containers filled with LMs rotate, and the LM surface assumes a
paraboloidal shape as a result. 
Stabilization or, more generally, control, would ``polish'' the liquid 
mirror from small surface waves due to seismic vibrations. 
Furthermore, feedback control \cite{ref11,Mirho2} could pave the way to
more advanced uses of LM  surfaces, including adaptive
optics. This could be extended to transparent conducting
fluids for refractive optics: “liquid lenses” already exist
\cite{ref16}, but they can only be adjusted in focal length. Also, they
suffer from aberrations, which could be electromagnetically controlled.

%==========================================================================
\section{Experimental setup: 
electromagnetic liquid metal pump and free-surface tile.}\label{SecSetup}
%==========================================================================
A Galinstan flow  was  sustained  by  an  
induction pump consisting  of a ferromagnetic rotor 
with permanent magnets mounted on it \cite{ref10}. The
magnetic field is partly ``frozen'' in  the liquid metal surrounding
the rotor. Therefore, as the field rotates, the liquid metal rotates
as well, although with a slip factor.  
This approach was preferred to 
conventional pumps, which would enter in electrical contact with the
metal flow, and could be damaged by intense currents.

\begin{figure}[!t]
\begin{center}
\includegraphics[width=\textwidth]{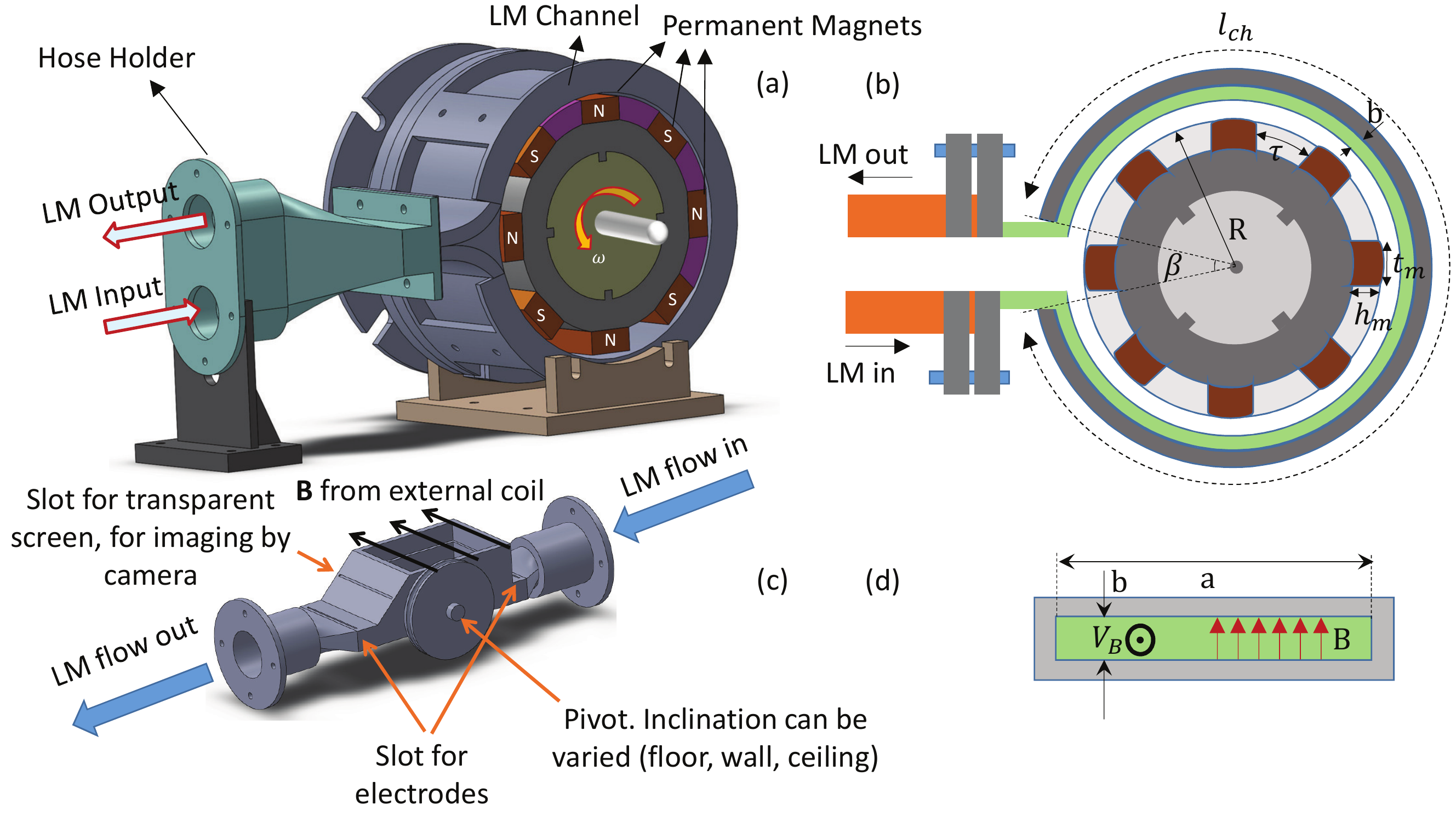}  %frv .eps}
\caption{a)Electromagnetic pump design. For the initial prototype, all the 
  parts shown (except the ferromagnetic core) were additively  manufactured by
  3D printing. b) Lateral and d) frontal cross-sections of the
  pump, with liquid metal shown in green. c)  Computer  rendering  of
  the  ``tile''  where  free-surface liquid  metal  flows  are  studied
  and  stabilized.  The  tile  can  be inclined to simulate the floor,
  vertical wall or ceiling of a fusion  reactor.}
\label{pump}
\end{center}
\end{figure}

A schematic and two cross-sections of the pump are shown in 
Fig.\ref{pump}a, b and d.  
The dimensions are: rotor radius $R$=9 cm, duct width $a$=7
cm, height of liquid metal channel $b$=3.5 mm, permanent magnets width
and height, $t_m$=2.54 cm and  $h_m$=1.27 cm, angle between inlet and
outlet $\beta$=30$^o$. 

\begin{figure}[!t]
\begin{center}
\includegraphics[width=0.8\textwidth]{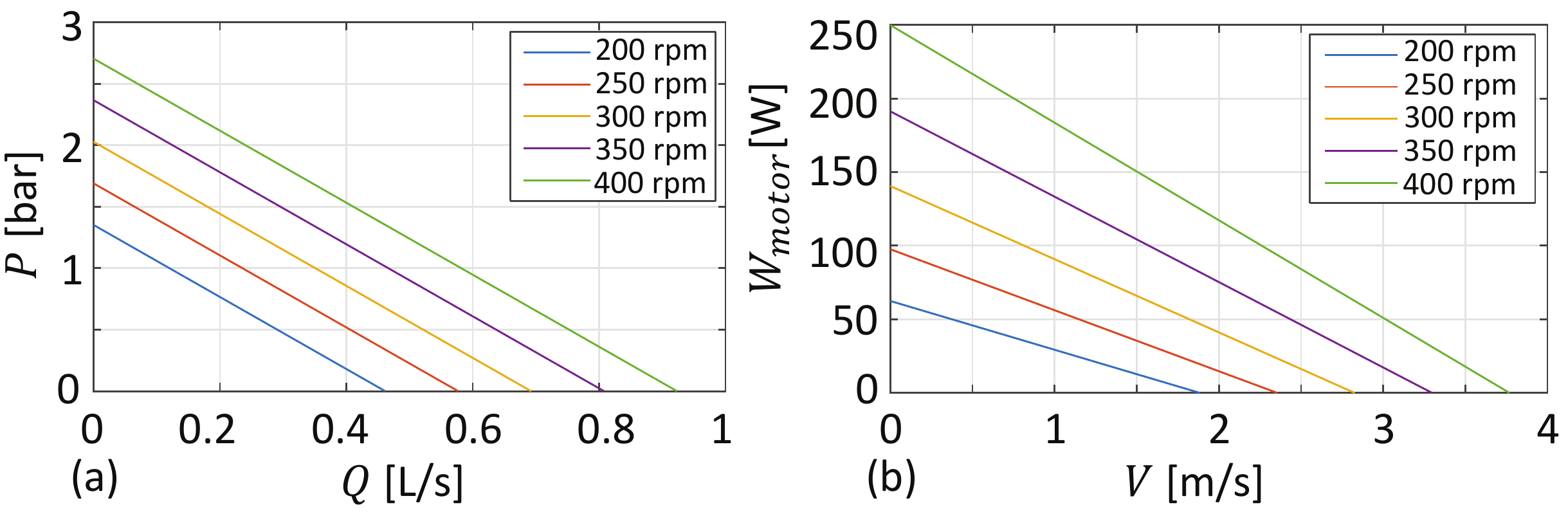}  %frv .eps}
\caption{(a) Pressure-flow rate and (b) motor power-fluid velocity
  characteristics of the pump for different angular
  frequencies of the rotor. 
  %Pump characteristics are: $R$ = 9 cm,
  %$a$ = 7 cm, $b$ = 3.5 mm, $t_m$= 2.54 cm, $h_m$= 1.27 cm,
  %$\beta$=$30^o$ and $\sigma=3.5\times10^6$  S/m.
}
\label{Pressure_Plot}
\end{center}
\end{figure}

The maximum pressure $P$ developed by the pump can be written as a function 
of the electrical conductivity $\sigma$, the slipping factor $s=1-V/V_B$ 
between the rotor velocity $V_B$ and fluid velocity $V$, the magnetic field 
$B$, mean length $l_{ch}$ of the MHD channel and negative transverse-end 
effect coefficient $k$ \cite{ref10}:
\begin{equation} 
  P=0.5\sigma s V_B B^2 l_{ch} k. 
\end{equation}
This pressure is plotted in Fig.\ref{Pressure_Plot}a as a function of the 
flow rate, 
\begin{equation} 
  Q=Vab, 
\end{equation}
where $a$ and $b$ are the transverse sizes of the MHD channel in 
Fig.\ref{pump}d. 

The product $PQ$, where $P$ and $Q$ are related to each other as in 
Fig.\ref{Pressure_Plot}a, represents the minimum motor power to rotate the 
rotor at a certain angular frequency $\omega \propto V_B$. 
This estimate, however, neglects Ohmic losses in the LM, 
\begin{equation} 
  P_{loss}=\frac{2P^2ab}{\sigma B^2l_{ch}k}. 
\end{equation}
The motor power needed can ultimately be calculated as $W_{motor}=PQ+P_{loss}$. 

Fig.\ref{Pressure_Plot}b illustrates the fluid velocities achievable by 
operating a motor of power $W_{motor}$ at higher and higher $\omega$. 
The initial prototype was nearly entirely realized in 3D-printed PLA 
plastic, to which Galinstan is not corrosive. 
With an electric 1 kW motor, it sustained Galinstan flows of up to 30 cm$^3$/s.

\begin{figure}[!t]
\begin{center}
\includegraphics[width=\textwidth]{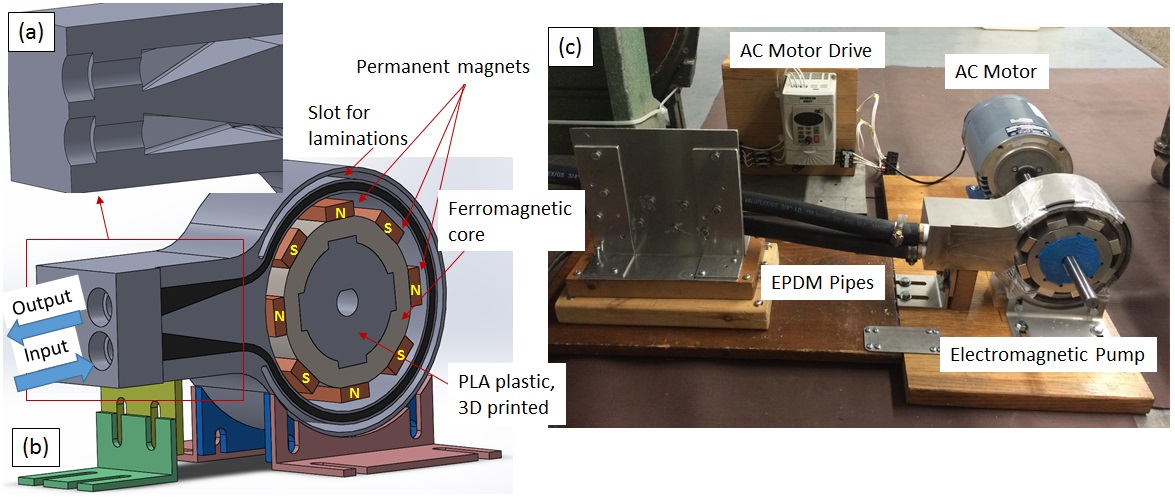}
\caption{More robust metallic pump with ferromagnetic laminations for 
increased pumping strength. a) Detail of CNC-machined taper from rectangular 
to circular cross-section, b) scheme of metallic pump and c) photo of 
LM pumping circuit}
\label{New_Pump}
\end{center}
\end{figure}

3D-printing was rapid and inexpensive. However, for increased mechanical 
strength, the pump was eventually rebuilt in non-magnetic stainless steel, 
with an improved design (Fig.\ref{New_Pump}).  
We started with a single block with a large bore  
closely fitting to the rotor. Then, from 
one side of the block a Computer-Numeric-Control (CNC) machine excavated 
a duct of uniform cross-sectional area but variable cross-sectional shape. 
The shape varied from a 69$\times$5 mm rectangle in proximity of the rotor, 
to take advantage of the strongest field, to  
18.5$\times$18.5 mm squares at the junctions with EPDM rubber hoses 
of circular cross-section. Finally, a CNC-machined lid 
(in black in Fig.\ref{New_Pump}b) was welded on the block (in gray) 
to laterally seal the channel. 

To increase $B$ and thus the pumping pressure $P$, 
ferromagnetic laminations were inserted in dedicated 
slots in the stator, concentric to the duct, at slightly outer radii 
(Fig.\ref{New_Pump}b). 
The lamination provided a low-reluctance path for 
the magnetic field lines through the MHD channel, which in turn
increased $P$ proportionally to $B^2$. The
latter reduced the negative effect of losses due to eddy currents
inside the body of the new metallic stator.

The hydraulic  circuit was  mostly ducted, but
comprised  a 3D-printed plastic, 5$\times$10 cm tile (Fig.\ref{pump}c)  for
free-surface  flows,  which typically are 0.5-1 cm thick. The tile was
tiltable  to  simulate the floor,  wall or ceiling of a fusion reactor, and
orientations in between. 

\begin{figure}[!t]
\begin{center}
\includegraphics[width=0.55\textwidth]{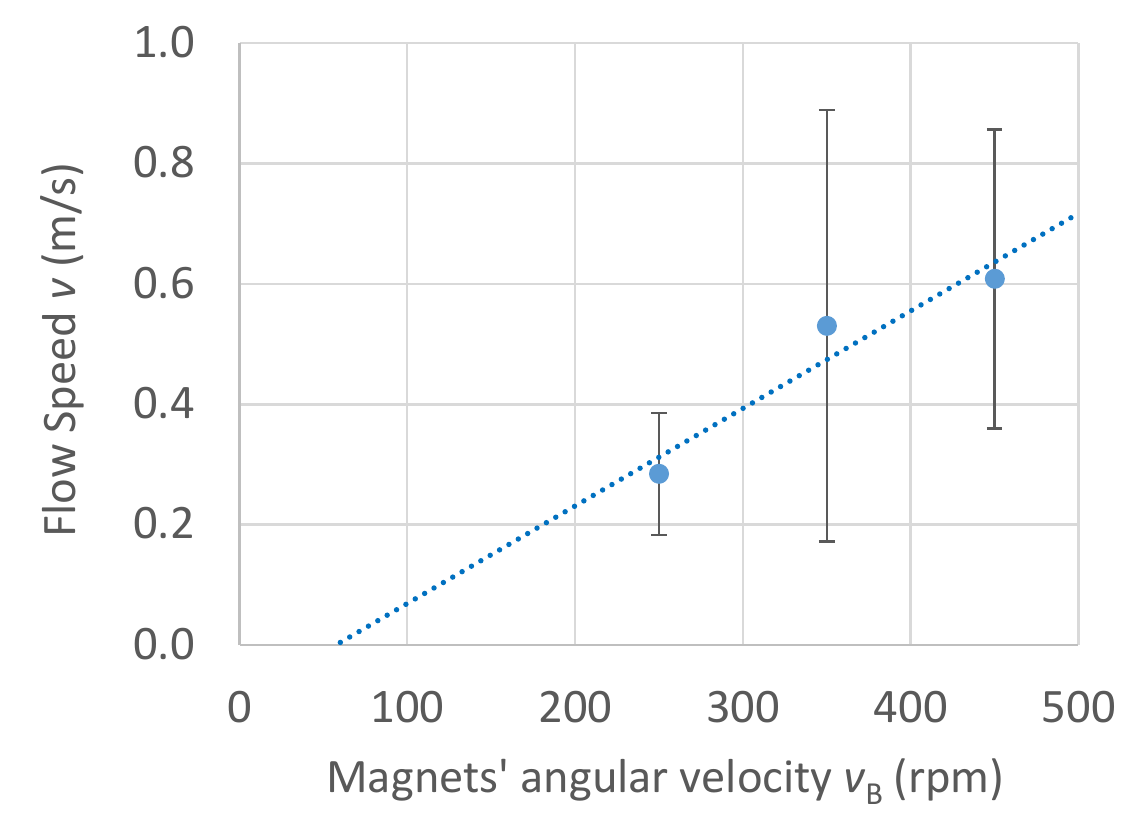}
\caption{Calibration of Galinstan flow velocity as function of permanent 
magnet rotor velocity.}
\label{FigCalibr}
\end{center}
\end{figure}

\subsection{Flow velocity calibration}
It was mentioned above that, due to the LM not being a perfect conductor, 
the field is not perfectly ``frozen-in'', and 
the LM flow slips with respect to the rotor. 
For this reason, it was important to calibrate the flow velocity $V$ 
as a function of the rotor velocity $V_B$. The flow velocity was measured 
by particle velocimetry, 
by means of fast-camera videos aquired at 2000 frames per second, and 
by means of photos of exposure time $t$=0.5-2 ms. 
From the videos and photos it was possible to infer the speed of grains of 
dust and other natural impurities dragged by the LM, and assume that that was 
also the LM speed. The results of the calibration are summarized in 
Fig.\ref{FigCalibr}. The large error bars are due to measuring the non-uniform, 
 time-evolving flow over several times and locations. 
On the other hand, the very fact that the flow is uneven and not constant makes 
it well-suited for stabilization experiments.

%==========================================================================
\section{Passive electromagnetic stabilization.}        \label{SecPass}
\begin{figure}[!t]
\begin{center}
  \includegraphics[width=0.57\textwidth]{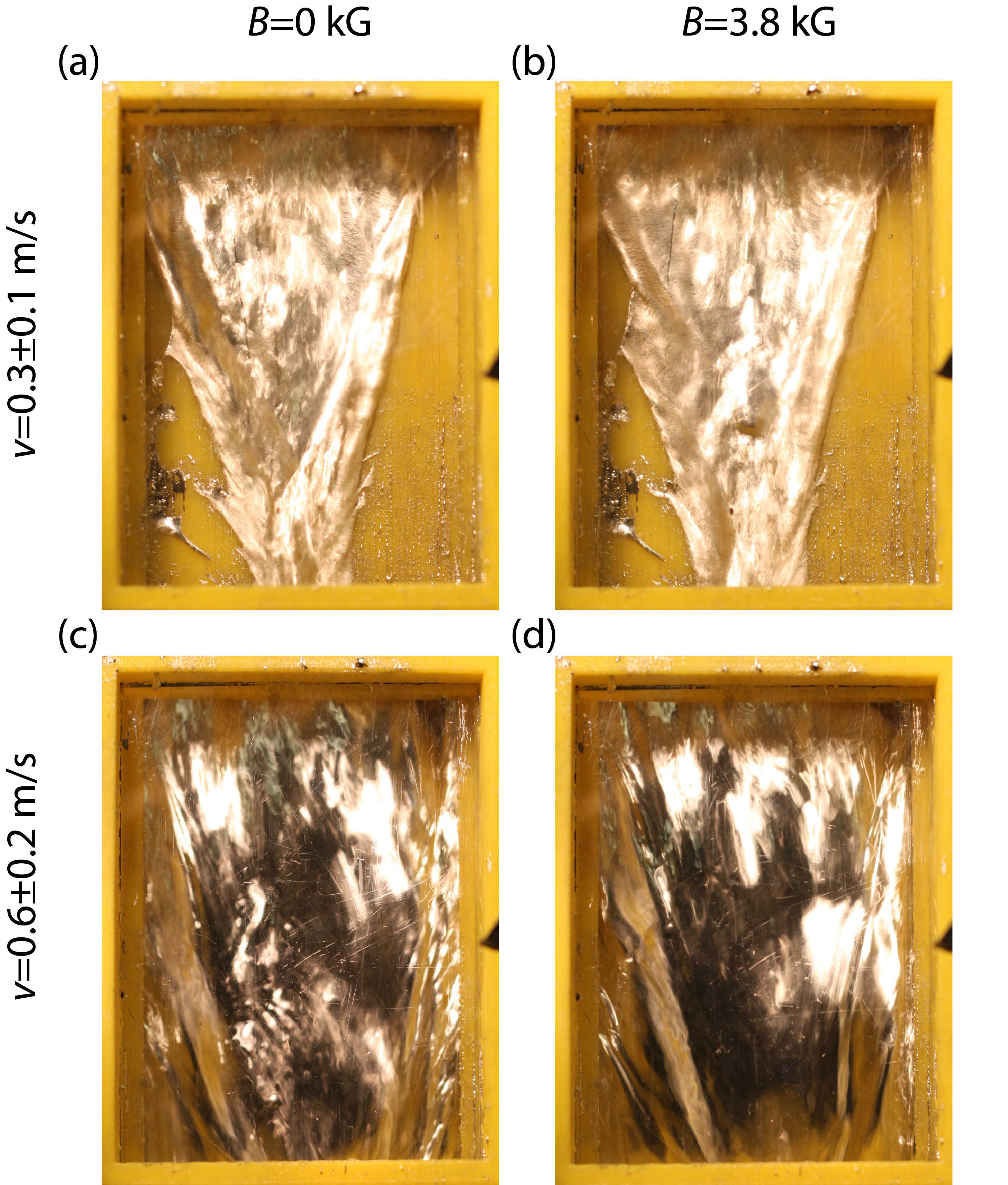}
  % convert -quality 50 -colorspace rgb PassStab_2x2_50_ImprLabels.jpg
  %                                     PassStab_2x2_50_ImprLabels_arxiv.jpg 
\caption{Liquid metal flow at an inclination $\theta$=35$^{\rm o}$ with respect 
to horizontal, for various velocities $v$ and transverse magnetic fields $B$. 
The photos 
illustrate the stabilizing effect of a high $B$, especially at 
high $v$, when the stabilizing force (scaling with $v$) is comparatively 
more important than surface tension (independent from $v$).}
\label{DC_Field_Effect}
\end{center}
\end{figure}

%==========================================================================
The tile shown in Fig.\ref{pump}c, with an approximately 5 mm thick layer 
of Galinstan flowing at velocity $v$,  
was placed in a transverse magnetic field $B$. 
A scan of $v$ up to 0.6$\pm$0.3 m/s and of $B$ up to 0.38 T 
provided the results shown in Fig.\ref{DC_Field_Effect}.  
In the absence of magnetic field, small amplitude
perturbations are visible over the surface of the flow. By applying
the magnetic field, the velocity fluctuations are strongly damped.

%==========================================================================
\subsection{Interpretation} 
%==========================================================================
The electromagnetic force term in the Navier-Stokes  equation 

\begin{equation}
\frac{\partial {\bf v}}{\partial t} +
({\bf v} \cdot \nabla){\bf v}=-\frac{{\bf \nabla} p}{\rho} 
+\nu\nabla^2{\bf v} + {\bf g} +
\frac{1}{\rho}({\bf j}\times{\bf B})
\label{Navier}
\end{equation}

contains,  through the  generalized Ohm's law

\begin{equation}
{\bf j}=\sigma({\bf E}+{\bf v}\times{\bf B})
\label{Ohm}
\end{equation}

a stabilizing term
$\frac{\sigma}{\rho}({\bf v}\times{\bf B})$ of order
$\sigma UB^2/ \rho$. This  term dominates over the non-linear
convective term  $({\bf v} \cdot \nabla){\bf v}$, of order
$U^2/L$: the ratio of the two, for  $B$=0.4  T, $U$ =0.2  m/s, $L$
=0.1  and  Galinstan conductivity  and  density,  evaluates approximately 44.  

In fact, the electromagnetic term also dominates over the viscous term
$\nu\nabla^2{\bf v}$: their ratio is the Hartmann  number
$Ha=BL\sqrt{\sigma/ \mu}$ ,  which  in  the  case  of interest is of
order  7$\cdot$10$^4$. 

%CHECK WHERE WE ARE IN SMOLENTSEV's PLOT

Therefore, in the limit of  no  applied  electric
field  and  negligible displacement  current,  the  Navier-Stokes
equation simplifies as follows: 

\begin{equation}
\frac{\partial {\bf v}}{\partial t} =
-\frac{{\bf \nabla} p}{\rho}+ {\bf g} +
\frac{\sigma}{\rho}({\bf v}\times{\bf B})\times{\bf B}
\label{Simplified_Navier}
\end{equation}

That is to say that motion is governed by  ${\bf \nabla} p $
(basically, the pump), by gravity and by a term proportional to 
$({\bf v}\times{\bf B})\times{\bf B}$. The latter 
is clearly stabilizing: if we define the $z$ axis as the magnetic
field direction, a fluctuation of velocity $\delta v_x$ in the $x$
direction results in a current density fluctuation 
$\delta j_y = - \sigma \, \delta v_x \, B$ and hence in a force per unit 
volume $\delta F_x /V= - \sigma \, \delta v_x \, B^2$. 
Note that this force is proportional to $\delta v_x$ and opposite to it, 
hence it acts as a viscous drag. Also note that it scales like $B^2$.  

Similarly, in the $y$
direction   $\delta F_y\propto -B^2\delta v_y$. Furthermore, by 
incompressibility the volumetric strain rate evaluates 
${\bf \nabla} \cdot {\bf v}=0$. As a consequence, 
if $\delta v_x$  and  $\delta v_y$  are kept small by stabilization,
then $\delta v_z$  is also small. 

%==========================================================================
\section{Active electromagnetic stabilization.}
\label{SecAct}
%==========================================================================
Flat electrodes described elsewhere \cite{ref11}  
were used to apply currents longitudinal to the flow  
and thus apply {\bf j}$\times${\bf B} forces orthogonal to the 
substrate. 

\begin{figure}[!t]
\begin{center}
  \includegraphics[width=0.8\textwidth]{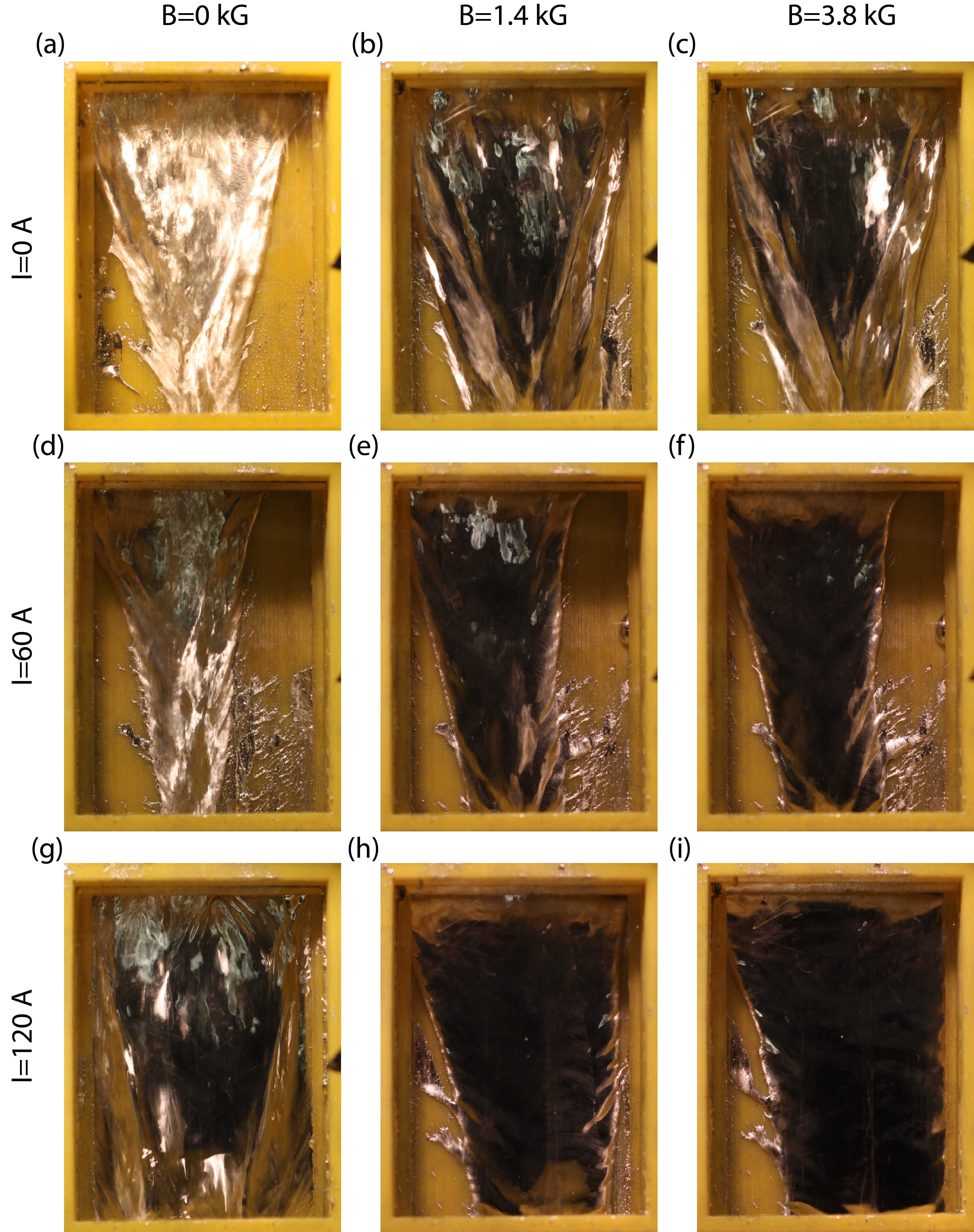}
  %convert -quality 50 -colorspace rgb ActStab_3x3_50_ImprLabels.jpg
  %                                    ActStab_3x3_50_ImprLabels_arxiv.jpg
\caption{Evidence of active stabilization of liquid metal flow at 
an inclination $\theta$=35$^{\rm o}$ with respect 
to horizontal, at fixed velocity $v=0.6\pm0.3$ m/s, for various currents $I$ 
(thus, current densities $j$) and transverse magnetic fields $B$. 
The photos illustrate the stabilizing effect of stronger and stronger 
${\bf j}\times {\bf B}$ force densities.}
\label{jxB}
\end{center}
\end{figure}

Fig. \ref{jxB} presents qualitative results obtained with Galinstan
flowing in the presence of $B$=0.38 T and $I$=0-120 A at 0.6$\pm$0.3  
m/s. Shown are nine images of the 5 cm wide free-surface flow, 
acquired for increasing values of $I$ and $B$. 

\begin{figure}[!t]
\begin{center}
  \includegraphics[width=0.85\textwidth]{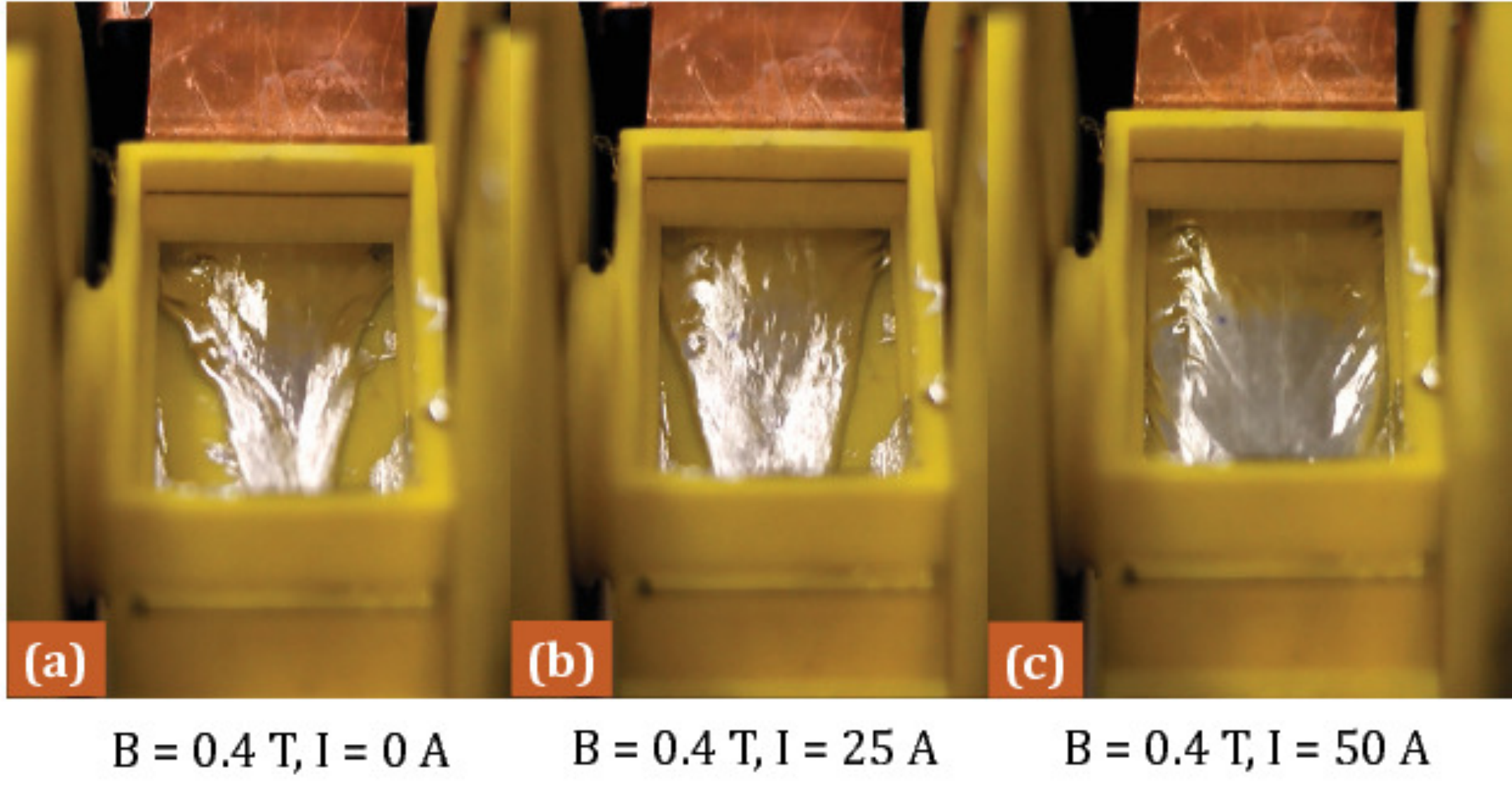}   %frv .eps}
  % convert -quality 100 -colorspace rgb jxB.eps jxB_arxiv.eps
\caption{Electromagnetically restrained flow in Galinstan. A scan of
  applied current at fixed magnetic field provides evidence that
  sufficient Lorentz force counteracts surface tension and other
  forces, and enables full coverage of an underlying solid wall.}
\label{jxB_old}
\end{center}
\end{figure}

In the ducted part of the tile, the LM flow has
rectangular cross-section. However, in the free-surface part of the
tile, with the constraints of one wall now removed, the cross-section
of the flow is free to change. Indeed, it does change in 
Fig.\ref{jxB}, as a consequence of
surface-tension trying to minimize the outer surface. In addition, 
gravity accelerates the 
flow and reduces its cross-section. In a reactor
these effects would leave part of the solid wall and the plasma exposed to 
each other. It is
reassuring, though, that when a sufficient Lorentz force is
applied in the normal direction, the flow flattens and covers the tile 
nearly entirely (Fig.\ref{jxB}i) or entirely (Fig.\ref{jxB_old}c).  
Full-coverage depends on velocity and inclination, and can sometimes be 
obtained at reduced ${\bf j}\times{\bf B}$ (Fig.\ref{jxB_old}c). 

In summary, a liquid metal flow 
in the presence of a DC magnetic field is stabilized if sufficiently 
intense transverse currents are applied.

\subsection{Future work} 
\begin{figure}[!t]
\begin{center}
\includegraphics[width=0.85\textwidth]{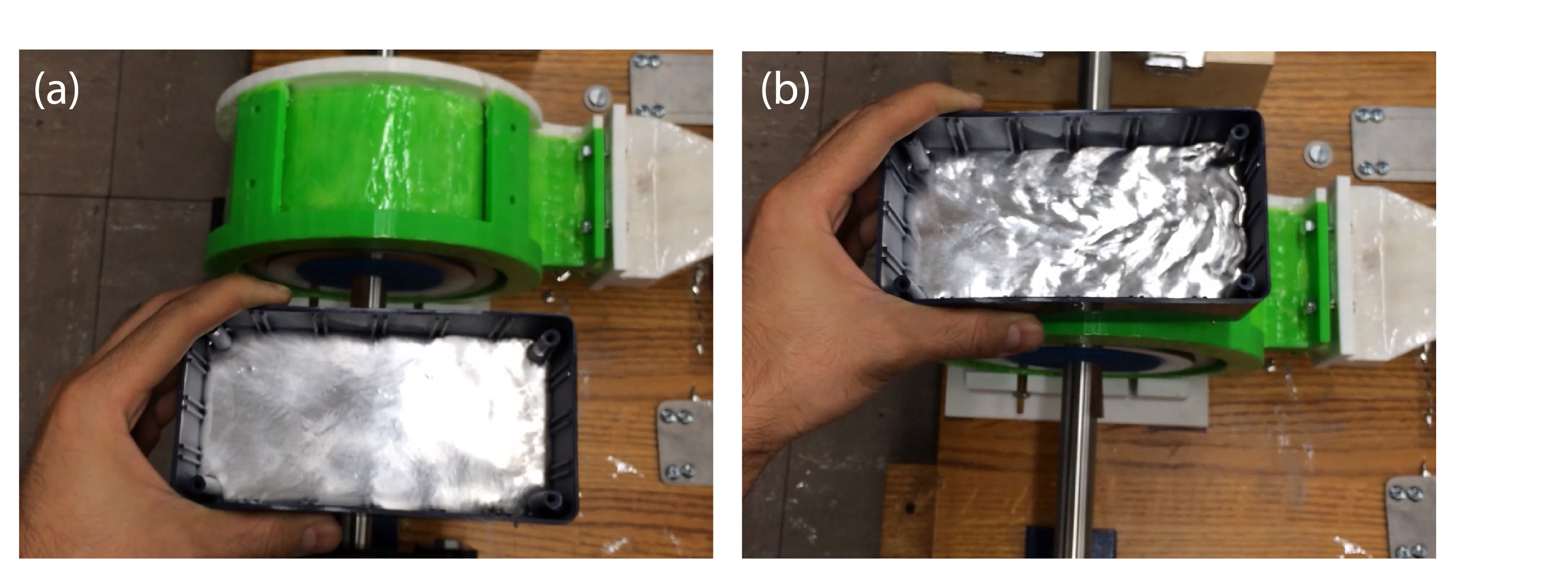}
\caption{Liquid metal is quiescent when (a) far from rotating permanent magnets 
of electromagnetic pump (in green) but (b) becomes uneven and turbulent in its 
proximity. Similar roughness could develop in the liquid walls of a reactor  
in the presence of rotating modes in the plasma or, in general, 
of non-uniform, time-evolving fields. Also note that 
dents in the side-walls of the 
container (in black) exacerbates the turbulence; dents and other realistic 
features will be present in the solid substrate of liquid walls in a reactor.}
\label{toy}
\end{center}
\end{figure}

In addition to the applied ${\bf j}$ and ${\bf B}$, 
undesired ${\bf j}$ and ${\bf B}$ are often present in the LM. 
The resulting undesired forces call for the need to adjust the applied 
${\bf j}$ in real time, in feedback with local LM thickness measurements 
\cite{ref11,Mirho2} . 
One such situation might occur when rotating instabilities -and the associated 
rotating helical field ${\bf B}$- are present in the 
plasma, inducing a helical ${\bf j}$ in the LM. 
A toy model of this phenomenon is easily reproduced in the 
laboratory by means of the rotating permanent magnets featured in the pump 
(Fig.\ref{toy}). 

A further motivation for feedback stems from the fact that the  
forces applied to the LM and pointing outward will never 
be perfectly uniform. The LM would thus be Rayleigh-Taylor unstable, similar 
to a pellet subject to inward-pointing forces in inertial confinement fusion 
\cite{Curreli}.  

Several other situations requiring feedback control were discussed in 
Ref.\cite{ref11}. 

%==========================================================================
\section{Summary and conclusions.}    \label{Sec6}
%==========================================================================

Results were presented about the electromagnetic stabilization 
of a free-surface flow of Galinstan. 
The flow was sustained by a rotating permanent magnet induction pump. 
Strong transverse magnetic fields ${\bf B}$ were found to stabilize the flow. 
This was due to the velocity fluctuations perpendicular to ${\bf B}$ 
causing a fluctuation of ${\bf j}$, through generalized Ohm's law. 
The resulting ${\bf j}\times{\bf B}$ force-density, scaling like $B^2$,  
opposes the original velocity fluctuation. Parallel fluctuations are also 
small, as a result of incompressibility. 

The external application of an additional, 
sufficiently high ${\bf j}$ orthogonal to ${\bf B}$ and parallel to the 
liquid metal layer resulted  
in ${\bf j} \times {\bf B}$ forces. Such forces 
pushed the free-surface flow against the substrate, acting like an  
``effective gravity'', also with stabilizing effects.

%-------------------- End of Body

%-------------------- References     

% -------------------- End of References

%\lastpageno	% This command sends the number of the last page to 
		% the MHD headline. Please latex your file twice if
		% you have used it.

%%%%%%%%%%%%%  
%	Place your tables and figures here
%%%%%%%%%%%%%

\end{document}